\documentstyle[mprocl,epsfig]{article}

\newcommand{\nubar}[0]{\overline{\nu}}
\newcommand{\numu}[0]{\nu_{\mu}}
\newcommand{\nuebar}[0]{\overline{\nu}_{e}}
\newcommand{\nue}[0]{\nu_{e}}
\newcommand{\numubar}[0]{\overline{\nu}_{\mu}}
\newcommand{\muebar}{(\numu \nuebar)}
\newcommand{\mubare}{(\numubar \nue)}

\newcommand{\Vcd}{\rm |V_{cd}|}
\newcommand{\Vcs}{\rm |V_{cs}|}
\newcommand{\Vub}{\rm |V_{ub}|}
\newcommand{\Vcb}{\rm |V_{cb}|}

\newcommand{\beq}{\begin{equation}}
\newcommand{\eeq}{\end{equation}}
\newcommand{\beqs}{\begin{eqnarray}}
\newcommand{\eeqs}{\end{eqnarray}}
\def\beq{\begin{equation}}
\def\eeq{\end{equation}}

\def\beq{\begin{equation}}   \def\eeq{\end{equation}}
 \newcommand{\gsim}{\lower.7ex\hbox{$
\;\stackrel{\textstyle>}{\sim}\;$}}
\newcommand{\lsim}{\lower.7ex\hbox{$
\;\stackrel{\textstyle<}{\sim}\;$}}

\def\lsim{\mathrel{\rlap{\lower3pt\hbox{\hskip0pt$\sim$}}
    \raise1pt\hbox{$<$}}}         
\def\gsim{\mathrel{\rlap{\lower4pt\hbox{\hskip1pt$\sim$}}
    \raise1pt\hbox{$>$}}}         
\newcommand{\stw}{\mbox{$\sin^2\theta_W$}}

\newcommand{\nubmu}{\overline{\nu_{\mu}}}
\newcommand{\nube}{\overline{\nu_{e}}}

\newcommand{\rmt}{\rm\textstyle}

\begin{document}

\pagestyle{plain}

\title{High Rate Physics at Neutrino Factories
\footnote{Presented at the 23rd Johns Hopkins Workshop on
Current Problems in Particle Theory, ``Neutrinos in the Next Millennium'',
Johns Hopkins University, Baltimore MD, June 10-12, 1999.}
}

\author{B.J. King}
\address{Brookhaven National Laboratory, Building 901A,
P.O. Box 5000, Upton, NY11973\\
email: bking@bnl.gov\\
web page: http://pubweb.bnl.gov/people/bking}

\maketitle

\abstracts{
 Both muon colliders and non-colliding muon storage rings using muon
collider technology have the potential to become the first
true ``neutrino factories'', with uniquely intense and precisely
characterized neutrino beams that could usher in a new era of high
rate and long baseline neutrino physics studies at accelerators.
This paper gives an overview of the predicted capabilities of
neutrino factories for high rate neutrino physics analyses that
will use huge event samples collected with novel, high performance
neutrino detectors.}

\section{Introduction}
\label{sec:intro}

 The idea of using muon storage rings for neutrino physics is an
old one~\cite{oldmusr}. More recent feasibility studies and design
work for muon colliders~\cite{snowmass,statusreport} led to
investigations~\cite{bjkphd,nufnal97} of the exciting neutrino
physics possibilities from the uniquely intense neutrino beams
they will produce and, as a variation on this theme that is now attracting
very considerable interest, it was then proposed~\cite{geer} to use muon
collider technology for non-colliding muon storage rings dedicated
to neutrino physics. An extensive literature~\cite{kirkoscref}
is now building up on the impressive potential of both types of
``neutrino factories'' (or ``nufacts'', for short).

 Nufacts will be used for each of the two classes of neutrino
experiments at accelerators:
\begin{enumerate}
  \item  high rate (HR) experiments, where
the detector is placed close to the neutrino source to obtain
the most intense beam possible and hence gather very high event
statistics of neutrino interactions.
  \item  long baseline (LB) experiments, where a very
massive neutrino detector is placed far away
from the neutrino source, deliberately sacrificing event
rate in order to study baseline-dependent properties of the
neutrinos and, in particular, whether there are ``flavor
oscillations'' in the types of neutrinos composing the
beam.
\end{enumerate} 

 The large muon currents and tight collimation of the neutrinos from
nufacts results in extremely intense beams with several important advantages
over the neutrino beams produced today from pion decay at accelerator
beamlines:
\begin{enumerate}
 \item  event statistics for HR experiments that
might be three or more orders-of-magnitude larger than in today's
HR neutrino experiments
 \item  both higher
statistics and longer baselines for LB experiments
 \item  extremely well understood and pure two-component beams with accurately
predictable energy spectra, angular divergences and intensities
 \item  the first high flux electron-neutrino and electron-antineutrino
beams at high energies.
\end{enumerate}

 Of the two classes of neutrino experiments at nufacts, LB neutrino
oscillation studies are currently attracting an enormous amount
of interest, with a large and growing literature~\cite{kirkoscref}
that includes another paper~\cite{Goodman} in these proceedings.
This paper will instead concentrate on the less-developed topic
of the potential for a rich and broad-based program of HR
neutrino physics at nufacts. It summarizes the material covered
in a much longer that is in preparation on this topic~\cite{numcbook}.

 The advantages of neutrino beams from the decays of stored
muons over traditional neutrino beams from pion decays are
in some ways even more notable for HR experiments than
for oscillation studies. In particular, the beam intensity and
uniquely small transverse extent close to production invites
the use of compact fully-active tracking targets backed by high-rate,
high-performance detectors.

 In contrast to long baseline neutrino oscillation measurements,
the physics interest lies in the interactions of neutrinos
rather than in their internal properties.
Neutrinos are unique in participating only in the weak
interaction and so, for example, they provide a probe of nucleon
structure that is
intrinsically cleaner than the alternative of charged
lepton (electron, positron or muon) scattering.
The weak interaction couplings of neutrinos to both quarks
and electrons are also interesting in their own right, as
will be discussed in the sections~\ref{sec:ew}
and~\ref{sec:ckm} on electroweak measurements and
the CKM quark mixing matrix, respectively. Nufacts will
also have potential for examining, or searching for, rare and
exotic interaction processes. As a bonus outside neutrino
physics, HR experiments at nufacts will be impressive factories
for the study of charm decays.

 The following section presents background material
on the expected experimental conditions for HR
neutrino physics at nufacts, in preparation for the
subsequent physics sections on the topics of nucleon
structure and QCD measurements, precision electroweak studies,
quark mixing measurements, rare and exotic processes and,
last but not least, charm decay physics, before the summary
section.

\section{Experimental Overview}
\label{sec:expt}

\subsection{The Neutrino Beam}
\label{ss:expt_beam}

 The beam spectrum from muon decays in a monochratic muon beam
is a completely pure 2-component mixture; the decays of parent $\mu^-$
provide beams of $\numu$ plus $\nuebar$ and $\mu^-$ decays provide
beams of $\numubar$ plus $\nue$:
\begin{eqnarray}
\mu^- & \rightarrow & \nu_\mu + \overline{\nu_{\rm e}} + {\rm e}^-,
                                             \nonumber \\
\mu^+ & \rightarrow & \overline{\nu_\mu} + \nu_{\rm e} + {\rm e}^+.
                                                 \label{eq:nuprod}
\end{eqnarray}
These beams will be denoted as $\muebar$ and $\mubare$ in
the rest of this paper. The kinematics of a muon decaying to
an electron and 2 neutrinos is precisely specified by the
electroweak theory and leads to precisely modeled neutrino
spectra for HR physics at nufacts.
This is a substantial advantage
over conventional neutrino beams from pion decays, particularly
for the high-statistics precision measurements that will be
described in sections~\ref{sec:sf} through~\ref{sec:ckm}.

 Reference~\cite{numcbook}
derives explicit expressions for the beam spectra at high
rate experiments in the context of a simplified but relatively
realistic model of a thin, monochromatic muon beam in the
production straight section of the storage ring.
The derivations assume zero net polarization over the data sample;
in order to minimize beam modeling systematics in experimental
analyses, it is likely that
the storage ring will be designed so that the muon polarization
precesses to average to zero over each fill of muons.

 The derivation in reference~\cite{numcbook} begins from the
neutrino energy distributions in the rest frame of the decaying
muon, $E_\nu'$, where the scaled energy, $x \equiv 2 E_\nu'/m_\mu$,
has a distribution given by
\begin{eqnarray}
\frac{dN_{\nu_\mu}}{dx} & = & 6.x^2 - 4.x^3
             \nonumber \\
\frac{dN_{\nu_e}}{dx} & = & 12.x^2 - 12.x^3
             \label{eq:beamx}
\end{eqnarray}
for muon-type neutrinos or anti-neutrinos
and for electron-type neutrinos or anti-neutrinos, respectively.
It is shown that $x$ is simply related to the neutrino energy
in the laboratory frame, $E_\nu$, by:
\begin{equation}
E_\nu(x,\theta') = x \frac{E_\mu}{2} \left( 1 + \beta \cos \theta' \right),
             \label{eq:beam_eboost}
\end{equation}
where $\beta = 1$ to a very good approximation and each angle in
the muon rest frame,
$\theta'$ corresponds to an angle $\theta$ in the laboratory
frame according to:
\begin{equation}
\sin \theta = \frac{\sin \theta'}{\gamma (1 + \beta \cos \theta')},
             \label{eq:beam_angboost}
\end{equation}
where $\gamma \equiv \frac{E_\mu}{m_\mu c^2}$.
Equation~\ref{eq:beam_eboost} combined with equations~\ref{eq:beamx}
and~\ref{eq:beam_angboost} therefore specifies the neturino energy
spectra at any position in an experimental target, at least in this
simplified model.

 Substituting the specific value $\theta'=\frac{\pi}{2}$
into equation~\ref{eq:beam_angboost}
shows that the forward hemisphere in the muon rest frame
is boosted into a narrow cone of half angle
\begin{equation}
\theta_\nu \simeq \sin \theta_\nu = 1/\gamma =
\frac{m_\mu c^2}{E_\mu} \simeq \frac{0.106}{E_\mu [{\rm GeV}]}.
                                                   \label{eq:thetanu}
\end{equation}
This characteristic opening half-angle
with respect to the muon beam direction will
include approximately half of the neutrinos in
the thin pencil beams for HR experiments
at nufacts and the radius subtended by this angle
at the experimental target is a reasonable choice for the
radius of the target. This corresponds, for
example to a 20 cm radius at 100 m from production in
a 50 GeV muon beam or at 1 km from a 500 GeV muon beam.

\subsection{Primer on Neutrino-Nucleon Deep Inelastic Scattering}
\label{ss:expt_dis}

 The dominant weak interaction processes for many-GeV neutrinos
and anti-neutrinos
are charged current (CC) and neutral current (NC) deep inelastic
scattering (DIS) off nucleons (N, i.e. protons and neutrons)
with the production of several hadrons ($X$):
\begin{eqnarray}
\nu (\overline{\nu}) + N & \rightarrow & \nu (\overline{\nu}) + X
          \;\;\;\;\;\;\;(NC)
                                        \nonumber \\
\nu + N & \rightarrow & l^- + X 
          \;\;\;\;\;\;\;(\nu-CC)       
                                        \nonumber \\
\overline{\nu} + N & \rightarrow & l^+ + X
          \;\;\;\;\;\;\;(\overline{\nu}-CC),
                                        \label{eq:nuint}
\end{eqnarray}
where the charged lepton, $l$, is an
electron/muon for electron/muon neutrinos.
The interaction of neutrinos with electrons is three orders of
magnitude less common than neutrino-nucleon DIS.
From the property
of lepton universality, the DIS interactions $\nue$'s and $\numu$'s,
or of their anti-neutrinos, are expected to be identical up to
final-state lepton mass corrections that are calculable and
almost negligible for multi-GeV neutrinos. This simplifies the
physics analyses and allows for a useful check of experimental
systematics through the comparision of DIS event samples from
the two neutrino flavors.
The CC (NC) DIS interactions are well described by the
``naive quark-parton model'' as quasi-elastic
(elastic) scattering off one of the many quarks (q) inside the
nucleon through the exchange of a virtual W (Z) boson:
\begin{eqnarray}
\nu (\overline{\nu}) + q & \rightarrow & \nu (\overline{\nu}) + q
          \;\;\;\;\;\;\;\;\;(NC)
                                        \label{eq:ncnuq} \\
\nu + q^{(-)} & \rightarrow & l^- + q^{(+)} 
          \;\;\;\;\;\;\;(\nu-CC)       
                                        \label{eq:ccnuq} \\
\overline{\nu} + q^{(+)} & \rightarrow & l^+ + q^{(-)}
          \;\;\;\;\;\;\;(\overline{\nu}-CC),
                                        \label{eq:ccnubarq}
\end{eqnarray}
To conserve charge, the initial-state and final-state
quarks in equations~\ref{eq:ccnuq}
and~\ref{eq:ccnubarq} differ by one unit of charge and have
been labelled accordingly, where
$q^{(-)} \in d,s,\overline{u},\overline{c}$ and
$q^{(+)} \in u,c,\overline{d},\overline{s}$.
The final state quark always ``hadronizes'' at the nuclear
distance scale, producing quark-antiquark pairs that arrange
into the several hadrons seen in the detector.

 The naive quark-parton model will be used in the rest of
this paper to give qualitative understanding of physics processes,
although it should be understood that the actual analyses will
require more sophisticated modeling.

\subsection{The Role of High Performance Neutrino Detectors}
\label{ss:expt_det}

\begin{figure}[t!] 
\centering
\includegraphics[height=3.5in,width=3.5in]{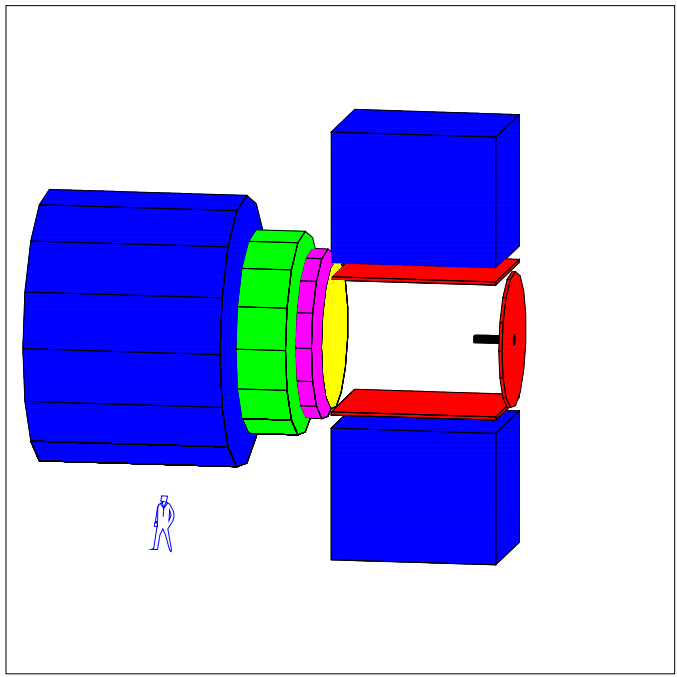}
\caption{Example of a general purpose neutrino detector. A human figure in
the lower left corner illustrates its size. The neutrino target is the small
horizontal cylinder at mid-height on the right hand side of the detector. Its
radial extent corresponds roughly to the radial spread of the neutrino pencil
beam, which is incident from the right hand side. Further details are given
in the text.}
\label{detector_fig}
\end{figure}

 The huge reduction in the neutrino beam cross section over the
conventional meters-wide beams from pion decay allows the use,
for the first time in neutrino physics, of
compact, specialized targets surrounded by high performance
detectors. As an example, figure~\ref{detector_fig} illustrates
the sort of HR general purpose neutrino detector that
would be well matched to the intense neutrino beams at nufacts.
(Subsection~\ref{ss:sf_pol}, on polarized targets, and
subsection~\ref{ss:ew_nue}, on neutrino-electron scattering
will discuss other possible examples of specialized
targets and detectors.)


 The neutrino target in figure~\ref{detector_fig} forms a notable 
contrast with the kiloton-scale calorimetric targets used in today's
HR neutrino experiments. Instead, it might comprise a 2 meter
long stack of CCD tracking planes with a radius of 20 cm chosen to
match the beam radius at approximately 100 meters from
production for a 50 GeV muon beam. As a detailed example, it might
contain 1500 planes of 300 micron thick silicon CCD's, corresponding to
a mass per unit area of approximately 100 ${\rm g.cm^{-2}}$, about
5 radiation lengths and one interaction length. (Further parameters
for this target will be presented in the following subsection.)

  Besides providing the mass for neutrino
interactions, a tracking target will allow precise reconstruction of the event
topologies from charged tracks, including event-by-event vertex tagging
of those events containing charm hadrons or tau leptons and, with the
higher energy beams, beauty hadrons.
Given the favorable vertexing geometry and the $\sim 3.5\;\mu$m
typical~\cite{SLD}
CCD hit resolutions, it is reasonable to expect~\cite{numcbook} perhaps
50 percent efficiency for charm tagging and ultrapure tagging for the
relatively rare beauty hadrons that will appear at the higher energies.

 The target in figure~\ref{detector_fig} is surrounded by a
time projection chamber (TPC) tracker in a vertical
dipole magnetic field. The
characteristic dE/dx signatures from the tracks would identify
each charged particle. Further particle ID is provided by the Cherenkov
photons that are produced in the TPC gas then reflected by a spherical
mirror at the downstream end of the tracker and focused onto a read-out
plane at the upstream end of the target.
The mirror is backed by electromagnetic and hadronic calorimeters
and, lastly, by iron-core toroidal magnets for muon ID.

 The relativistically invariant quantities that are routinely
extracted in DIS experiments, along with their interpretations
in the naive quark-parton model, are 1) Feynman $x$, the fraction
of the nucleon 4-momentum carried by the struck quark, 2) the
inelasticity, $y = E_{\rm hadronic}/E_\nu$,
which is related to the scattering angle of the neutrino in the
neutrino-quark CoM frame (where $E_{\rm hadronic}$ is the energy contained
in the hadron shower), and 3) the momentum-transfer-squared,
$Q^2 = 2 M_p E_\nu x y$ for $M_p$ the mass of the proton.
The high performance detectors at
nufacts will have the further capability
of reconstructing the hadronic 4-vector, resulting in a much better
characterization of each interaction, particularly for NC interactions.

\subsection{Cross Sections and Event Rates}
\label{ss:expt_xsec}

 This section derives approximate estimates of the event sample
sizes expected in HR targets at nufacts.

  The DIS cross sections for multi-GeV neutrinos are approximately
proportional to the neutrino energy, $E_\nu$, and the charged current
(CC) and neutral current (NC) interaction cross sections for neutrinos
and antineutrinos have numerical values of~\cite{quigg}:
\begin{equation}
 {\rm \sigma_{\nu N}\; for\;}
 \left(
 \begin{array}{c}
   \nu-CC \\
   \nu-NC \\
   \overline{\nu}-CC \\
   \overline{\nu}-NC
 \end{array}
 \right)\;
 \simeq
 \left(
 \begin{array}{c}
    0.72 \\ 0.23 \\ 0.38 \\ 0.13
   \end{array}
 \right)
\times {\rm E_\nu [GeV]}
\times 10^{-38}\: {\rm cm^2}.
                                            \label{eq:xsec}
\end{equation}

 This energy dependence allows us to express the number of DIS
events in a target as the simple product of the
average beam energy, $<E_\nu>$, and the integrated luminosity:
\begin{equation}
{\rm No.\; DIS\: events} = {\rm \sigma^R_{\nu N}[cm^2.GeV^{-1}]} \times
<E_\nu>[{\rm GeV}] \times \int L {\rm dt}\; [{\rm cm^{-2}}]
\label{eq:eventno}
\end{equation}
where the units are given in square brackets and the
constant of proportionality is the reduced cross section
defined by:
\begin{equation}
{\rm \sigma^R_{\nu N}} \equiv {\rm \sigma_{\nu N}} / E_\nu.
\end{equation}
Summing equation~\ref{eq:xsec} over NC and CC interactions and
averaging over neutrinos and anti-neutrinos gives the numerical
value of
\begin{equation}
{\rm \sigma^R_{\nu N}} = 0.73 \times 10^{-38}\: {\rm cm^2.GeV^{-1}}.
       \label{eq:sigmaRnuN}
\end{equation}

  For a target subtending $\theta_\nu = 1/\gamma$ from the
production straight section, the number of neutrinos passing
through the target is seen to be exactly equal to the number of
muon decays, $N^{ss}_\mu$, since each muon decay produces two
neutrinos but only half of them will be in the forward hemisphere
in the muon rest frame. In this case, the total integrated
luminosity into the detector is clearly
\begin{equation}
\int L {\rm dt}\; [{\rm cm^{-2}}] =
                       6.022 \times 10^{23} \times N^{ss}_\mu
                       \times l[{\rm g.cm^{-2}}],
\end{equation}
where $6.022 \times 10^{23}$ is Avogadro's number and
$l[{\rm g.cm^{-2}}]$ is the detector mass per unit area
in appropriate units.

\begin{table}[htb!]
\begin{center}
\caption{Specifications, integrated luminosities and event rates 
for the HR targets discussed in this paper and for
50 GeV (500 GeV) muon storage rings. The
approximation is made that the target is situated 100 m (1 km)
downstream from a straight section that has $N^{ss}_\mu=10^{20}$
decays of 50 GeV (500 GeV) muons. This corresponds to average
neutrino energies of 32.5 GeV (325 GeV) and to approximately
1 (2) years running for storage ring parameters given previously
in the literature.}
\begin{tabular}{|r|ccc|}
\hline
target purpose & general  & polarized  & $\nu-e$ scatt. \\
\hline
material       & Si CCD's & solid HD   & liquid ${\rm CH_4}$ \\
ave. density   & 0.5 ${\rm g.cm^{-3}}$  & 0.267 ${\rm g.cm^{-3}}$
               & 0.717 ${\rm g.cm^{-3}}$ \\
length         & 2 m      &  0.5 m     &  20 m \\
mass/area, $l$ & 100 ${\rm g.cm^{-2}}$ 
               & 13.4 ${\rm g.cm^{-2}}$
               & 1434 ${\rm g.cm^{-2}}$ \\
radius         & 0.2 m    &  0.2 m     & 0.2 m \\
mass           & 126 kg   &  16.8 kg   & 1800 kg \\ 
$\int L {\rm dt}$
               & $6.0 \times 10^{45}\;{\rm cm^{-2}}$
               & $8.1 \times 10^{44}\;{\rm cm^{-2}}$
               & $8.6 \times 10^{46}\;{\rm cm^{-2}}$ \\
no. DIS events:&&&\\
at 50 GeV      & $1.4 \times 10^9$
               & $1.9 \times 10^8$
               & $2.0 \times 10^{10}$\\
at 500 GeV     & $1.4 \times 10^{10}$
               & $1.9 \times 10^9$
               & $2.0 \times 10^{11}$\\
no. $\nu$-e events:&&&\\
at 50 GeV      & $3.5 \times 10^5$
               & NA
               & $7 \times 10^{6}$\\
at 500 GeV      & $3.5 \times 10^6$
               & NA
               & $7 \times 10^{7}$\\
\hline
\end{tabular}
\label{tab:events}
\end{center}
\end{table}

Table~\ref{tab:events} gives a summary of some characteristics
for examples of the three types of targets discussed in this paper,
and also gives realistic but very approximate integrated luminosities
and event sample sizes for 2 illustrative nufact energies:
50 GeV and 500 GeV. A muon beam energy of about 50 GeV is a likely
choice for a dedicated muon storage ring~\cite{Lyon}, with default
specifications of $10^{20}$ muon decays per year in the straight
section providing the neutrino beams. Five hundred GeV muons
corresponds to a 1 TeV center-of-mass muon collider and this
parameter set is discussed in reference~\cite{numcbook}>

The event samples in table~\ref{tab:events}
are truly impressive. It is seen, for example, that
high performance detectors with fully-active tracking
neutrino targets might collect and precisely reconstruct
data samples with many billions of neutrino-nucleon DIS
interactions -- more than three orders of magnitude larger
than any of the data samples collected using today's much
larger and cruder neutrino targets.

\section{Nucleon Structure Functions and QCD Studies}
\label{sec:sf}

\subsection{Structure Function Measurements with Unpolarized Targets}
\label{ss:sf_unpol}

  It can be shown quite generally that the differential cross sections
for neutrino-nucleon and anti-neutrino interactions can be written
in terms of nucleon {\em structure functions} (SF), $F_1$, $F_2$ and $F_3$,
as:
\begin{equation} \label{crsec}
\frac{d^2 \sigma^{\nu (\bar \nu)}}{dx dy} = \frac{G^2_F M_p E_\nu}{\pi}
\left[ x y^2 F_1^{\nu (\bar \nu)} +
\left( 1- y \right) F_2^{\nu (\bar \nu)}
\mp xy \left( 1-y/2 \right ) F_3^{\nu (\bar \nu)}
\right],
\end{equation}
with $G_F$ the Fermi couplng constant, $M_p$ the proton mass and
$x$ and $y$ in~\ref{ss:expt_det} and where small correction factors
have been neglected for simplicity.
The SF are found to exhibit approximate scaling behavior for
$Q^2 \to \infty$, i.e. $F_i (x, Q^2) \to F_i (x)$,
which is an encouraging approximate verification of the exact
scaling behaviour that is predicted in the naive quark-parton model.

The SF must be experimentally determined
by measuring the differential cross sections as functions of $x$, $y$
and $Q^2$ and extracting the SF using binned fits that exploit their
differing $y$ dependences.
The parity-violating structure function, $xF_3$, can only be measured
in neutrino-nucleon scattering (and at HERA, although much less
precisely and in a different
kinematic regime). The $F_1$ and $F_2$ SF's that
are defined for neutrino-nucleon scattering also probe different
combinations of quarks to the analagous SF's defined for charged
lepton DIS experiments.

 Superscripts on the SF in equation~\ref{crsec}
distinguish the potentially different structure functions for $\nu$
and $\bar \nu$ scattering. However, neutrino experiments to date on
isoscalar targets have struggled to resolve the differences, due to
insufficient statistics and insufficient control of experimental systematics.
Nufacts will therefore provide the first clean extractions of
all 6 SF's and this will provide an opportunity that is unique,
for any physics process, to lay bare the quark-by-quark content
of the nucleon. This will now be made plausible using a simplified
discussion in the context of the naive quark-parton model.

  In the framework of the naive quark-parton model, the SF
can be expressed in terms of quark densities as:
\begin{eqnarray}
2F_1^{\nu N}(x,Q^2) &=& d(x) + s(x) + \bar u(x) + \bar c(x)
\nonumber \\
2F_1^{\bar \nu N}(x,Q^2) &=& u(x) + c(x) + \bar d(x) + \bar s(x)
\nonumber \\
F_3^{\nu N}(x,Q^2) &=& d(x) + s(x) - \bar u(x) - \bar c(x)
\nonumber \\
F_3^{\bar \nu N}(x,Q^2) &=& u(x) + c(x) - \bar d(x) - \bar s(x)
\label{eq:sfquarks}
\end{eqnarray}
and
\begin{equation}
F_2(x,Q^2)  \equiv  2xF_1(x,Q^2).
\end{equation}
An isoscalar target (i.e. equal numbers of neutrons and
protons, as is the case for the silicon tracking target of
subsection~\ref{ss:expt_det}) is expected to provide four
further approximate equalities:
\begin{eqnarray}
u(x) &=& d(x) \nonumber \\
\bar u(x) &=& \bar d(x) \nonumber \\
s(x) &=& \bar s(x) \nonumber \\
c(x) &=& \bar c(x). \label{eq:iso}
\end{eqnarray}
The 8 equations in~\ref{eq:sfquarks} and~\ref{eq:iso} can
therefore be solved in this naive model to derive, from
the measured SF's, the quark-by-quark density distributions,
at each $x$, for isoscalar targets. (In detail, the first
two equations of~\ref{eq:sfquarks} are actually identical
in this simplified model, so additional information from,
e.g., charm tagging is needed to completely solve for all
of the quark densities.)

 As an extension of this argument, SF measurements from both
hydrogen and deuterium targets would, using isospin symmetry,
give measurements of the quark-by-quark densities in both protons
and neutrons. Thus, even the polarization-averaged measurements
from the\ polarized hydrogen targets in subsection~\ref{ss:sf_pol}
will clearly be extremely valuable.

 In real life, there will clearly be complications beyond
the simple picture from the naive quark model, including
nuclear effects, so-called ``higher twist'' processes and
charm thresholds, to name just a few. Despite the complications,
the above discussion amounts to at least a plausibility
argument that nufacts will, for the first time
in any experimental process, have the potential to
disentangle the quark-by-quark structure of isoscalar
nuclear targets, protons and neutrons.

 The uniquely precise and detailed quark-by-quark
characterization of nucleon structure will provide an invaluable
reference source for many diverse analyses in collider and fixed target
physics including, of course, the other precision analyses at
nufacts. Precise measurements at high $x$, $x \rightarrow 1$,
are particularly relevant to the modeling of rates for interesting
physics processes and backgrounds at hadron colliders.

 The universality and applicability of the quark-parton model
will be able to be further tested by comparing the nufact
SF's with those from charged lepton DIS measurements, where
the quarks couple to the electromagnetic
current in proportion to their charge squared.
As an
example, the quark model relationship between the $F_2$ structure 
functions is
\begin{equation}
F_2^\mu = \frac{5}{18} F_2^\nu + \frac{1}{6}x(s(x) + {\bar s}(x)).
\end{equation}
Current global SF analyses find puzzling deviations from this relationship
at low $x$ which, it is speculated, might be related to differing nuclear
shadowing effects and would vanish for simple nucleon targets.
A first high-statistics measurement of $F_2^\nu$ on deuterium
at a nufact that can be compared with existing $F_2^\mu$
measurements on deuterium might well resolve the discrepancy.

 As the second broad area for SF studies at nufacts, DIS has long
been a dominant process in testing and understanding perturbative
quantum chromodynamics (pQCD). For example,
the scaling behavior of the $xF_3$ SF is broken by a predictable
pQCD evolution that is logarithmic in $Q^2$ and is independent,
to first order in perturbation theory, of the nucleon's gluon distribution.
The observed evolution of $xF_3$ in $\nu$N scattering already provides
one of the
most precise measurements of the strong coupling constant, $\alpha_s$,
as does its integral over $x$:
\begin{equation} \label{glssr}
\int^1_0 F_3(x, Q^2) dx =
3 \left[ 1 - \frac{\alpha_s}{\pi} + O\left(\frac{\alpha_s}{\pi} \right)^2
\right ],
\end{equation}
a measurement known as the
Gross Llewellyn-Smith (GLS) Sum Rule.
Both of these measurements of $\alpha_s$ are currently limited by
experimental systematic uncertainties and great improvements can
be expected with the superior experimental conditions at nufacts.

 Heavy quark production in neutrino interactions will provide both
additional complications and oppportunities for pQCD studies,
due to the additional mass scale, $m_Q$, for $Q=c,b$.
Nufacts will provide a unique facility
to test and extend theoretical treatments on this topic. Improved
understanding will be valuable for pQCD analyses both within
and beyond neutrino physics and, for example, correct modeling
of the heavy quark threshold suppression is vital for the
CKM studies of section~\ref{sec:ckm}.

 In addition to these inclusive charm production studies,
semi-inclusive measurements involving, for example,
the production of $\Lambda_{C}^{+}$ or $J/\psi$'s will provide
opportunities for further insights into QCD, as is addressed
elsewhere in these proceedings~\cite{Petrovproc}.

\subsection{Measurements with Polarized Targets}
\label{ss:sf_pol}

 As well as vastly improving on today's $\nu$N SF measurements
with unpolarized targets, nufacts will provide the first neutrino
beams with sufficient intensity to allow the use of polarized targets.
Neutrino scattering experiments using polarized targets have
considerable potential for further resolution of the structure of
the nucleon and for additional tests of QCD, even at the lower energy
beams of dedicated nufacts.

 Polarized lepton-nucleon DIS studies have so far been the domain of
charged lepton experiments, where a rich program
includes SLAC E155x, the recently approved COMPASS
experiment at CERN, HERMES at DESY,
ELSA in Bonn, MAMI in Mainz and experiments at the Thomas Jefferson National
Accelerator Facility. Related studies at collider energies will
soon become available in polarized proton-proton collisions at
BNL's RHIC collider, and HERA may also eventually polarize their
protons beam to provide polarized positron-proton collisions
at center-of-momentum energies of approximately 300 GeV.

 Polarized neutrino-nucleon scattering retains the experimental
advantages over charged lepton DIS experiments that were discussed
in subsection~\ref{ss:sf_unpol} for non-polarized targets. In addition, the
absence of significant target heating from the beam will allow the use
of polarized solid protium-deuterium (HD) targets that cannot survive
in charged lepton experiments and have so far only been used in
experiments with low intensity neutron or photon~\cite{sphice} beams.
The preparation of such targets is a detailed craft~\cite{Crabb & Meyer}
involving doping the targets with ortho-hydrogen and holding
them for long periods of time at very low temperatures and high
magnetic fields, e.g. 30-40 days at 17 T and 15 mK.
In order to avoid building an entire new detector around the
polarized target, it would be ecomonical to
place the polarized target directly upstream from another
detector, such as the tracking detector described in
subsection~\ref{ss:expt_det}.

 The polarized SF that might be measured at a nufact
are the transversely-polarized and longitudinally-polarized
parity-conserving spin-SF's, $g_1$ and $g_2$, respectively, as
well as the parity-violating spin-SF $g_5$.

 As was discussed above for the unpolarized parity-violating SF,
$xF_3$, polarized target experiments at nufacts should provide
easily the most precise measurements of the parity-violating
spin structure functions, $g_5$, that can only be
measured in CC weak interactions.
The only other future opportunity to measure $g_5$ that
has been widely discussed is the possibility of eventually
polarizing the proton beams in the HERA e-p collider.
 Because of kinematic constraints on reconstructing
events, a polarized HERA would be able to make less precise
measurements~\cite{HERApol} for protons 
in the complementary high $Q^2$ region,
$Q^2 > 225\:{\rm GeV}^2$, that will not be accessible
to nufacts. It would not provide measurements for
neutrons, of course.

 The quark content for the polarized SF's in
the naive quark model is as follows:
\begin{eqnarray}
g_1^{\nu N} &=&
\Delta d + \Delta s + \Delta\overline{u} + \Delta\overline{c}
\nonumber \\
g_1^{\bar \nu N} &=&
\Delta u + \Delta c + \Delta \overline{d} + \Delta \overline{s}, 
\nonumber \\
g_2^{\nu N} &=& 0  \nonumber \\
g_2^{\bar \nu N} &=& 0  \nonumber \\
g_5^{\nu N} &=&
\Delta d + \Delta s - \Delta\overline{u} - \Delta\overline{c}
\nonumber \\
g_5^{\bar \nu N} &=&
\Delta u + \Delta c - \Delta \overline{d} - \Delta \overline{s}, 
\label{eq:polsf}
\end{eqnarray}
where each
$\Delta q \equiv q^{\uparrow \uparrow} - q^{\uparrow \downarrow}$
is the difference between quarks polarized parallel to the nucleon spin
and those polarized anti-parallel.
By similar arguments to that presented in
subsection~\ref{ss:sf_unpol} for unpolarized SF's it can
be seen that the measurement of the polarized SF's
for both protons and neutrons should provide much
information about the quark-by-quark spin content of the
nucleon. This should give nufacts a central role in
resolving the current so-called ``spin crisis'' that
has been a dominant topic for DIS spin experiments over
the last decade, as follows.

 The spin crisis refers to the experimental
observation~\cite{spin_crisis} that only a small fraction of
the nucleon spin is contributed
by the quarks, which was considered to be counter-intuitive
and has lead to efforts to investigate the other possible
contributions to the spin. The contributions are
summarized in the helicity sum rule for the nucleon's
longitudinal spin, $S_z^N$:
\begin{equation}
S_z^N = \frac{1}{2} =
\frac{1}{2}(\Delta u + \Delta d + \Delta s) +
L_q + \Delta G + L_G,
\end{equation}
where the quark contribution is
$\Delta \Sigma = \Delta u + \Delta d + \Delta s$,
$\Delta G$ is the gluon spin and $L_q$ and $L_G$
are the possible angular momentum contributions
from the quarks and gluons circulating in the nucleon.

 The above method for extracting the various quark spin
distributions at nufacts, from inclusive SF, should be
much cleaner theoretically than the semi-inclusive measurements
needed in charged lepton experiments,
which rely on semi-inclusive measurements plus assumptions about
fragmentation functions. Even so, nufacts should also provide
novel and extended capabilities for such semi-inclusive measurements.
They can use, for example, the semi-muonic tagging of charm production.
This can be calibrated by vertex-tagging experiments in other
detectors and is sensitive to the spin of the strange quarks
in the nucleon and, perhaps in some kinematic regions, to the
spin contribution of the gluon. Such a capability, if realized,
would be very valuable in solving the spin crisis, particularly
since the gluon contribution is extremely hard to measure and
yet it is the leading suspect for providing the bulk of the
nucleon's spin.

\section{Precision Electroweak Studies}
\label{sec:ew}

  Nufacts will improve enormously on current neutrino experiments in
allowing access to aspects of the electroweak interaction that cannot be
readily probed at colliders. As the most important single electroweak
topic, they might well restore the historical role of neutrino scattering
experiments in providing some of the most precise measurements
of the weak mixing angle, $\stw$.
The electroweak theory predicts that $\stw$
is related to the mass ratio of
the W and Z intermediate vector bosons: 
\begin{equation}
\sin^2\theta_W = 1 - \left( \frac{M_W}{M_Z} \right) ^2
                        \label{eq:wma}
\end{equation}
to first order in perturbation theory.
Precise measurements in various processes can probe higher order
diagrams to test the
consistency of the electroweak theory and provide differing sensitivities
to new physics processes occurring,
for example, through loop diagram contributions to the scattering
amplitudes.

 Nufacts should provide vastly improved determinations of the
weak mixing angle, $\stw$, in measurements
from (i) the ratio of neutral current (NC) to charged current (CC)
DIS events and (ii) measuring absolute cross sections for
neutrino-electron scattering, as discussed in the following
subsections.

\subsection{Neutrino-Electron Scattering}
\label{ss:ew_nue}

 Neutrino-electron elastic scattering,
\begin{equation}
\nu e^- \rightarrow \nu e^-,
\end{equation}
is an interaction between point elementary particles with
a precise theoretical prediction for its cross section
as a function of $\stw$. It therefore provides measurements of
$\stw$ that will be essentially
limited only by statistics (3 orders of magnitude down from DIS)
and by ingenuity in minimizing the experimental uncertainties.

 Four different $\nu$-e elastic scattering processes occur in total
in the $\muebar$ and $\mubare$ beams:
\begin{eqnarray}
\numu e^-\to\numu e^- \label{reac:numu-e}\\
\nube e^-\to\nube e^- \label{reac:nube-e}\\
\nubmu e^-\to\nubmu e^- \label{reac:nubmu-e}\\
\nue e^-\to\nue e^-, \label{reac:nue-e}
\end{eqnarray}
where the first two processes occur in the $\muebar$ beam and
the final two in the $\mubare$ beam. The two scattering processes
in a given beam cannot be experimentally separated so the
experimental measurement involves counting the sum of events
from the two processes. The $\muebar$ and $\mubare$ beams will
provide two physically
distinct measurements because the scattering involves different
diagrams for the neutrino species in the two beams:
an s-channel (annihilation) diagram contributes to
equation~\ref{reac:nube-e} while
equation~\ref{reac:nue-e} includes a charged current
($W^\pm$ exchange) t-channel diagram.

 Because of the small ratio of the electron to proton mass, the
cross section for neutrino-electron scattering is much smaller than
that for DIS. The numerical values for the cross sections
after integrating over y are:
\begin{equation}
\sigma(\nu e^-\to\nu e^-) = 1.6 \times 10^{-41} \times E_\nu[GeV]
                            \times \left[ g_L^2+\frac{1}{3}g_R^2\right],
\label{eqn:nue-sigmaval}
\end{equation}
where the left-handed and right-handed coupling constants,
$g_L$ and $g_R$, are different for each of the processes in
equations~\ref{reac:nubmu-e} through~\ref{reac:nube-e}.
Their values and the values for the term in square brackets
are given in table~\ref{tab:glgr}.

\begin{table}\begin{center}
\begin{tabular}{|c|c|c|c|}\hline
Reaction & $g_L$ & $g_R$ & $g_L^2+\frac{1}{3}g_R^2$ \\ \hline
$\numu e^-\to\numu e^-$ & $-\frac{1}{2}+\stw$ & $\stw$ & 0.0925 \\ 
$\nube e^-\to\nube e^-$ & $\stw$ & $\frac  {1}{2}+\stw$ & 0.2258 \\
$\nubmu e^-\to\nubmu e^-$ & $\stw$ & $-\frac{1}{2}+\stw$ & 0.0758 \\ 
$\nue e^-\to\nue e^-$ & $\frac{1}{2}+\stw$ & $\stw$ & 0.5425 \\ 
\hline
\end{tabular}
\end{center}
\label{tab:glgr}
\caption{$g_L$ and $g_R$ by $\nu-e$ scattering process listed for the
cross section formula in equation~\ref{eqn:nue-sigmaval}. The value
$\stw=0.225$ has been used for the final column.}
\end{table}

The experimental signature for $\nu-e$ scattering is a single
negatively charged electron with very low transverse momentum,
$p_t\stackrel{<}{\sim}\sqrt{m_eE_\nu}$. A tracking detector with
very good $p_t$ resolution is needed to resolve the signal peak
from the much broader background distributions from quasi-elastic $\nu-N$
scattering and other low-multiplicity $\nu-N$ scattering events.
An attractive target/detector option is a
low-Z liquid that can form tracks of ionization electrons and
drift them to an electronic read-out. Liquids under
consideration~\cite{numcbook} include argon and methane or
other saturated alkanes, and the readout goemetry might
be a TPC or, to reduce pile-up backgrounds, a printed-circuit
kaptan strip
geometry~\cite{Rehak} with more channels and shorter drift distances.

 Besides background rejection, the other big experimental challenge
for the measurement will be the determination of the absolute
neutrino flux. For the $\muebar$ beam, signal processes can
probably~\cite{numcbook}
be precisely normalized to the theoretically predictable processes
involving muon production off electrons:
$\numu e^-\to\nue \mu^-$ and $\nube e^-\to\nubmu\mu^-$. The
$\mubare$ beam requires an additional stage of
relative flux normalization, which might be accomplished~\cite{numcbook}
using the relative
sizes of the event samples for quasi-elastic neutrino-nucleon
scattering, ${\rm \nu N \to  l^\pm N'}$, in the $\mubare$ and $\muebar$
runs.

 Specific example parameters for the target/detector and for the
event sample sizes are given in table~\ref{tab:events}.
Predicted event samples are in the range of millions to
tens-of-millions of events and if the experimental uncertainties
can be controlled then these sample sizes will correspond~\cite{numcbook}
to limiting statistical uncertainties of $\Delta \stw = 0.000\,3$
and $0.000\,1$ for the $\muebar$ and $\mubare$ beams, respectively,
at the 50 GeV nufact and $\Delta \stw = 0.000\,1$ and
$0.000\,03$ for the corresponding measurements
at 500 GeV. It remains, of course, to be demonstrated that the
experimental uncertainties will ever allow these measurement
accuracies, but the potential of this measurement is anyway
impressive.

\subsection{Measurement of the WMA in DIS}
\label{ss:ew_dis}

 The most precise current measurement of $\stw$ from
$\nu$-N DIS, from NuTeV~\cite{nutevwma},
\begin{equation}
\stw=0.2253\pm0.0019{\rmt (stat.)}\pm0.0010{\rmt (syst.)}
\;\;\;\;\;\;\;\;\;{\rm (preliminary)},
\end{equation}
gives an equivalent uncertainty on the W mass,
${\rm\Delta M_W \simeq 100\;MeV/c^2}$,
that is competitive with direct measurements at colliders.
Measurements at nufacts may well extend
the historical tradition of neutrino-nucleon DIS experiments
in providing some of the most precise measurements of $\stw$.

 The complexity of nucleon targets makes it necessary,
as in previous neutrino experiments, to consider NC-to-CC cross
section ratios in order to make theoretical sense of the results.
Since both the $\muebar$ and $\mubare$ beams at nufacts are mixtures
of a neutrino and and antineutrino flavor, the appropriate experimental
NC-to-CC ratios, $R_{\nu}^{\mu^{-}}$ and $R_{\nu}^{\mu^{+}}$, respectively,
are each linear combinations of the traditional NC-to-CC cross
section ratios for neutrinos, $R^\nu$, and antineutrinos,
$R^{\overline{\nu}}$:
$$
R_{\nu}^{\mu^{-}} \simeq 0.70 R^\nu + 0.30 R^{\overline{\nu}},
      \label{eq:Rmumlc}
$$
$$
R_{\overline{\nu}}^{\mu^{+}} \simeq 0.63 R^\nu + 0.37 R^{\overline{\nu}}.
      \label{eq:Rmuplc}
$$
Because the linear combinations are so similar,
the two measurements will have almost equal numerical values,
$R_{\nu}^{\mu^{-}} \simeq 0.330$ and
$R_{\overline{\nu}}^{\mu^{+}} \simeq 0.332$ for $\stw=0.225$,
and their physics content will clearly be nearly identical.

 Traditional heavy target neutrino detectors could not distinguish 
$\nu_{e}$-induced CC interactions from NC interactions and this
separation will probably be the most demanding experimental
requirement for analyses at nufacts. However, with huge statistics
and high performance tracking detectors such as those described
in section ??, the DIS measurement will presumably eventually be
systematically limited by theoretical hadronic uncertainties
rather than statistical or experimental uncertainties.

 The vastly improved statistics and experimental conditions
at nufacts makes it difficult to extrapolate the measurement
accuracy from that at today's neutrino experiments.
Reference~\cite{nufnal97} estimates that the predicted
uncertainty in $M_W$ from a nufact analysis might be of order
10 MeV, which improves by an order of magnitude on today's
neutrino experiments~\cite{nutevwma,ccfrwma} and is approximately
equal to the projected best direct measurements from future
collider experiments. Thus, the $\stw$ measurement from DIS
could be a useful complement to the $\nu$-e scattering measurements
described in the previous subsection.

\section{Measurements of CKM Matrix Elements}
\label{sec:ckm}

  With huge samples of flavor tagged events, nufacts have the
potential to make impressive measurements of the
absolute squares of several of the elements in the
fundamental Cabbibo-Kobayashi-Maskawa (CKM) mixing matrix
that characterises the charged current (CC) weak interactions of
quarks. Lower energy nufacts will provide an opportunity for
unique and precise measurements of the elements $\Vcd$ and $\Vcs$
and further measurements of the more theoretically interesting
elements $\Vub$ and $\Vcb$ will become available at the higher energies
required for B production. Each of the analyses~\cite{numcbook}
would involve vertex tagging of heavy final state quarks and will
be somewhat analagous to, but vastly superior to, current neutrino
measurements of ${\rm |V_{cd}|^2}$ that use dimuon events for
final state tagging of charm quarks.

 Today's measurements of $\Vcd$ in $\nu$N scattering are already
the most precise in any process and a fundamental advantage of
$\nu$N DIS over all other types of CKM measurements is that the
scattering process involves the interaction of an external
W boson probing the quarks inside a nucleon rather than an internal
W interaction inside a hadron. (In principle, the HERA ep collider
could also do such measurements, but they turn out not to be feasible
in practice.) Such external W probes allow
measurements that are theoretically cleaner than other processes
because the asymptotic freedom property of QCD predicts quasi-free
quarks with reduced influence from their hadronic environment for
$Q$ significantly above the GeV-scale.

 Another difference relative to, say, the CKM measurements at
B factories is that the measurements are of the magnitudes of
individual CKM matrix elements rather than of interference terms.
This is understood easily from the naive quark-parton model
approximation, where the differential cross sections for
quark transitions, $\frac{d\sigma}{dx}$,
are given as products of quark densities with the absolute 
squares of matrix elements:
\begin{eqnarray}
\frac{d\sigma}{dx}( d\rightarrow c) \propto x\, d(x)|V_{cd}|^2
\label{eq:Vcd} \\
\frac{d\sigma}{dx}( s\rightarrow c) \propto x\,s(x)|V_{cs}|^2
\label{eq:Vcs} \\
\frac{d\sigma}{dx}( u\rightarrow b) \propto x\,u(x)|V_{ub}|^2
\label{eq:Vub} \\
\frac{d\sigma}{dx}( c\rightarrow b) \propto x\,c(x)|V_{cb}|^2,
\label{eq:Vcb}
\end{eqnarray}
with $d(x)$, $s(x)$, $u(x)$ and $c(x)$ the respective initial-state
quark densities and these expressions are very approximate only
because important threshold correction factors have been
neglected.

 The $\Vcd$ analysis at a nufact, with hundreds of millions of
vertex-tagged charm events in a high-performance detector, 
should have profound statistical, experimental and theoretical
advantages over today's measurement and may well reach~\cite{numcbook}
the parts-per-mil level of accuracy.
As is evident from equation~\ref{eq:Vcs}, the measurement of $\Vcs$
is intrinsically more difficult than $\Vcd$ because it requires a
knowledge of the strangeness content of the nucleon.
Reference~\cite{numcbook} speculates that
$\Vcs$ could nevertheless be measured at the percent level,
which would also become the best direct measurement of the element
and would provide a much improved unitarity test on its value.

 The B-production analyses at higher energy nufacts should be
experimentally rather similar to the charm analyses but would
have vastly greater theoretical interest. Both $\Vub$ and $\Vcb$
determine the lengths of sides of the ``unitarity triangle''
that is predicted to exist if the CKM matrix is indeed unitary.
The main goal of today's B factories is to measure the
interior angles of this triangle to confirm that it is
indeed a triangle, and the complementary input from a nufact
will be an enormous help in this verification process. In
particular, the predicted~\cite{nufnal97,numcbook} 1-2 \%
accuracy in $\Vub^2$ is several times better than predicted
accuracies in any future measurements of other processes, and will
obviously provide a very strong constraint on the unitarity
triangle.

 Table~\ref{ckm_table} summarizes the predicted~\cite{numcbook} nufact
contributions to determining the CKM matrix elements, giving
the current, experimentally determined values for the 9 mixing
probabilities along with their current
percentage uncertainties and speculative projections~\cite{nufnal97} 
for how 4 of the 9 uncertainties might be reduced at a 500 GeV
nufact.

\begin{table}[ht!]
\caption{Absolute squares of the elements in the
Cabbibo-Kobayashi-Maskawa (CKM) quark mixing matrix.
The second row for each quark gives
current percentage uncertainties in the absolute squares
and speculative projections of the uncertainties after analyses
from a 500 GeV nufact. The measurements of $\Vcd^2$ and $\Vcs^2$
might be comparably good for a 50 GeV nufact but
 $\Vub^2$ and $\Vcb^2$ would not be measured.
The uncertainties assume that no unitarity
constraints have been used.}
\begin{tabular}{|c|lll|}
\hline
          & \hspace{0.2 cm} \bf{d} & \hspace{0.2 cm} \bf{s}  &
                                          \hspace{0.3 cm}\bf{b}  \\
\hline
\bf{u}    &   \bf{0.95}  &  \bf{0.05}    &  \bf{0.00001}  \\
          &   $\pm$0.1\%  &  $\pm$1.6\%    & $\pm$50\% $\rightarrow$ 1-2\% \\
          &&& \\
\bf{c}    &   \bf{0.05}  &  \bf{0.95}    &  \bf{0.002}  \\
          &   $\pm$15\% $\rightarrow$ 0.2-0.5\%   &
              $\pm$35\% $\rightarrow$ $\sim 1$\%         &
              $\pm$15\% $\rightarrow$ 3-5\%     \\
          &&& \\
\bf{t}    &  \bf{0.0001}  &  \bf{0.001}  &  \bf{1.0} \\
          &   $\pm$25\%   &  $\pm$40\%   & $\pm$30\% \\
\hline
\label{ckm_table}
\end{tabular}
\end{table}

\section{Rare and Exotic Processes}
\label{sec:rare}

 The potential for studying rare and exotic processes at
nufacts is limited relative to collider experiments due to the
more modest center-of-mass energies. Nevertheless
there will still be some opportunities to both
further our knowledge of the SM by studying rare processes
and to search for processes not predicted by the SM.
Examples of a few potential processes involving exotic
physics that might be observable at nufacts include:
\begin{enumerate}
  \item a non-standard flavor-changing neutral current (FCNC) interaction
converting valence u quarks to charm quarks, $u \rightarrow c$,
would appear as an excess of charm events over anti-charm events
at high $x$. In contrast to FCNC searches in decays, this is
a unique opportunity to observe fundamental
flavor-changing neutral currents at the quark level~\cite{numcbook}
  \item some types of unstable exotic neutral
leptons~\cite{numcbook}
  \item R-parity violating SUSY searches~\cite{deJongh}
  \item hypothetical new neutrino interactions involving microscopic
extra dimensions that are much larger than the Planck scale~\cite{deJongh}.
\end{enumerate}

\section{Charm Physics Studies}
\label{sec:charm}

  It should be clear from the discussions in
sections~\ref{ss:expt_det} and~\ref{sec:ckm} that
nufact's will be rather impressive factories for the study of charm -- with
a clean, well reconstructed sample of several times $10^8$ charmed hadrons
produced in $10^{10}$ neutrino interactions.

There are several interesting physics motivations for charm
studies at a nufact~\cite{numcbook}. Measurement of charm decay branching
ratios and lifetimes are useful for both QCD studies and for the theoretical
calibration of the physics analyses on B hadrons. Charm decays also
provide a ``clean laboratory'' to search for exotic physics contributions
since the SM predicts 1) tiny branching ratios for rare decays,
2) small CP asymmetries and 3) slow
${\rm D^0 \rightarrow \overline{D^0}}$ oscillations, with
only of order 1 in $10^4$ oscillating before decay.

 It is a unique advantage of CC-induced charm production in
neutrinos that the production sign of the charm quark is tagged
with very high efficiency and purity by the charge of the final
state lepton:
\begin{eqnarray}
\nu q \rightarrow l^-c \nonumber \\
\nubar\: \overline{q} \rightarrow l^+\overline{c},
\end{eqnarray}
where $q=d$ or $s$ and $\overline{q}=\overline{d}$ or $\overline{s}$.
This is of particular benefit to oscillation and CP studies,
as is the expected precise vertexing reconstruction of the
proper lifetime of decays. As an example of the advances in
charm studies that might result, particle-antiparticle
mixing has yet to be observed in the charm sector and 
it is quite plausible~\cite{nufnal97}
that a nufact would provide the first observation of
${\rm D^0 - \overline{D^0}}$ mixing.

\section{Summary}
\label{sec:sum}

 It has been shown that nufacts will have unique capabilities
for HR neutrino physics, with huge event samples collected
in high-performance detectors. The physics reach should extend
well beyond traditional neutrino physics topics and should
complement or improve upon many analyses in diverse areas
of collider and fixed target physics, including:
\begin{itemize}
 \item  the only realistic opportunity, in any physics process,
to determine the detailed quark-by-quark structure of the nucleon
 \item  with polarized targets, additionally map out the quark-by-quark
spin structure of the nucleon and, perhaps, determine the gluon contribution
to the nucleon's spin
 \item  some of the most precise measurements and tests of
perturbative QCD
 \item  some of the most precise tests of the electroweak theory
through measurements of $\stw$ with fractional uncertainties
approaching the $10^{-4}$ level
 \item  measurements of the CKM quark mixing matrix that will
be interesting for lower energy nufacts ($\Vcd$ and $\Vcs$)
and will become extremely important ($\Vub$ and $\Vcb$) at
higher energies
 \item  a new realm to search for exotic physics processes
 \item  as a bonus outside neutrino physics, a charm factory
with unique capabilities.
\end{itemize}

 The expected experimental conditions at nufacts are so novel
and impressive that the topics presented in this paper must
surely represent no more than a limited first attempt to
understand their full potential for HR neutrino
physics.

\section*{Acknowledgments}

This work was performed under the auspices of
the U.S. Department of Energy under contract no. DE-AC02-98CH10886.


\begin{thebibliography}{99}

\bibitem{oldmusr} An unpublished note by A.C. Melissinos (1960) is the
first reference found by K.T. McDonald in a compilation of 
muon collider accelerator physics prior to 1995 that can be found at
{\it http://www.hep.princeton.edu/mumu/physics/index.html\#accel}.
\bibitem{snowmass}  The Muon Collider Collaboration,
{\it $\mu^+\mu^-$ Collider: A Feasibility Study}, BNL-52503,
Fermilab-Conf-96/092, LBNL-38946, July 1996, unpublished.
\bibitem{statusreport}
 The Muon Collider Collaboration,
{\it Status of Muon Collider Research
and Development and Future Plans}, Phys. Rev. ST Accel. Beams, 3 August,
1999.
\bibitem{bjkphd}
     B.J. King,
    {\it Assessment of the prospects for muon colliders},
    paper submitted in partial fulfillment of requirements
    for Ph.D., Columbia University, New York (1994), available
    from LANL preprint archive as {\it physics/9907026}.
\bibitem{nufnal97}
B.J. King,
  {\it Neutrino Physics at a Muon Collider},
 Proc. Workshop on Physics at the First Muon Collider
 and Front End of a Muon Collider, Fermilab, November 6-9, 1997,
{\it hep-ex/9907033}.

\bibitem{geer} S. Geer,
 {\it Neutrino Beams from Muon Storage Rings: Characteristics
and Physics Potential}, Phys. Rev. {\bf D57}, 6989 (1998).
\bibitem{kirkoscref} A compilation of papers with physics topics
relevant to neutrino factories, compiled by K.T. McDonald, can be found at
{\it http://www.hep.princeton.edu/mumu/nuphys/index.html} . 
\bibitem{Goodman} M. Goodman, these proceedings.
\bibitem{numcbook} I. Bigi {\it et al.},
 {\it The Potential for High Rate Neutrino
Physics at Muon Colliders and Other Muon Storage Rings},
contact author B.J. King, to be published.
\bibitem{SLD} K. Abe {\it et al.}, {\it Design and performance
of the SLD vertex detector: a 307 Mpixel tracking system.},
NIM A 400 (1997) 287-343.
\bibitem{quigg} See, for example, Chris Quigg,
    {\it Neutrino Interaction Cross Sections}, FERMILAB-Conf-97/158-T.
\bibitem{Lyon} NuFact'99,~Lyon,~5-9~July,~1999,
{\it http://lyoinfo.in2p3.fr/nufact99/general.html}.
\bibitem{Petrovproc} A. Petrov, these proceedings.
\bibitem{sphice} D. Babusci {\it et al.},
  Proc. 11th Int. Symp. High-Energy Spin Phys., AIP Conf.
Proc. 343, p.523 (1995).
\bibitem{Crabb & Meyer} D.G. Crabb and W. Meyer,
 {\it Solid Polarized Targets for Nuclear and Particle Physics
Experiments}, Ann. Rev. Nucl. Part. Sci. 1997, 47:67-109.
\bibitem{HERApol} See, for example, A. Deshpande,
{\it The Physics Case for Polarized Protons at HERA}, hep-ex/9908051.
\bibitem{spin_crisis} J. Ashman {\it et al.}, the EMC Collaboration,
 Phys. Lett. B {\bf 206} (1988) 364.
\bibitem{Rehak} Private communication with P. Rehak.
\bibitem{nutevwma}
 K.S.McFarland, NuTeV collaboration, to be published
in the proceedings of the XXXIIIrd Rencontres de Moriond (1998).
\bibitem{ccfrwma}
C.Arroyo {\it et. al.}, Phys. Rev. Lett. {\bf 72}, 3452 (1994).
\bibitem{deJongh} Private communication with F. deJongh.


\end{thebibliography}
\end{document}